\documentclass[a4paper,12pt]{article} 

\usepackage[centertags]{amsmath}
\usepackage{amsmath}
\usepackage[dutch,british]{babel}
\usepackage{amsfonts}
\usepackage{amssymb}
\usepackage{amsthm}
\usepackage{newlfont}
\usepackage{color}
\usepackage{subfig}
\usepackage{bbm}
\usepackage{fancyhdr}
\usepackage{graphicx}
\usepackage{fancybox}
\usepackage{geometry}
\usepackage{psfrag}
\usepackage{multirow}
\usepackage{authblk}

\theoremstyle{plain}
  \newtheorem{theorem}{Theorem}[section]

  \newtheorem{lemma}[theorem]{Lemma}
  
\theoremstyle{definition}

\theoremstyle{remark}

\numberwithin{equation}{section}

\newcommand\otimesal{\mathop{\hbox{\raise 1.6 ex
  \hbox{$\scriptscriptstyle\mathrm{al}$}
\kern -0.92 em \hbox{$\otimes$}}}}
\newcommand\oplusal{\mathop{\hbox{\raise 1.6 ex
  \hbox{$\scriptscriptstyle\mathrm{al}$}
\kern -0.92 em \hbox{$\oplus$}}}}
\newcommand\Gammal{\hbox{\raise 1.7 ex
\hbox{$\scriptscriptstyle\mathrm{al}$}\kern -0.50 em $\Gamma$}}
\renewcommand\i{\mathrm{i}}

  \let\ga=\gamma 
\let\ka=\kappa

  \let\La=\Lambda

\newcommand{\caB}{{\mathcal B}}

\newcommand{\caH}{{\mathcal H}}

\newcommand{\caO}{{\mathcal O}}
\newcommand{\caP}{{\mathcal P}}

\newcommand{\bbA}{{\mathbb A}}

\newcommand{\bbC}{{\mathbb C}}

\newcommand{\bbN}{{\mathbb N}}

\newcommand{\bbR}{{\mathbb R}}

\newcommand{\bbZ}{{\mathbb Z}}

\newcommand{\opunit}{\text{1}\kern-0.22em\text{l}}

\newcommand{\e}{{\mathrm e}}

\renewcommand{\d}{{\mathrm d}}

\newcommand{\beq}{ \begin{equation} }
\newcommand{\eeq}{ \end{equation} }
\newcommand{\bet}{ \begin{theorem} }
\newcommand{\eet}{ \end{theorem} }

\newcommand{\baq}{\begin{eqnarray}}
\newcommand{\eaq}{\end{eqnarray}}

 \newcounter{smallarabics}
\newenvironment{arabicenumerate}
{\begin{list}{{\normalfont\textrm{\arabic{smallarabics})}}}
  {\usecounter{smallarabics}\setlength{\itemindent}{0cm}
  \setlength{\leftmargin}{5ex}\setlength{\labelwidth}{4ex}
  \setlength{\topsep}{0.75\parsep}\setlength{\partopsep}{0ex}
   \setlength{\itemsep}{0ex}}}
{\end{list}}

\newcounter{smallroman}

\newcommand{\ben}{\begin{arabicenumerate}}
\newcommand{\een}{\end{arabicenumerate}}

\newcommand{\norm}{ \|}
\newcommand{\str}{ |}

\newcommand{\adjoint}{\mathrm{ad}}
\newcommand{\ad}{\adjoint}

\newcommand{\su}{(\tfrac{1}{\nu})}

\newcommand{\union}{P}

\begin{document}

\title{A rigorous theory of many-body prethermalization for periodically driven and closed quantum systems}

\author[1]{Dmitry Abanin}
\author[2]{Wojciech De Roeck}
\author[1]{Wen Wei Ho}
\author[3]{Fran\c{c}ois Huveneers}
\affil[1]{Department of Theoretical Physics, University of Geneva, Switzerland}
\affil[2]{Instituut voor Theoretische Fysica, KU Leuven, Belgium}
\affil[3]{CEREMADE, Universit\'e Paris-Dauphine, France}
\renewcommand\Authands{ and }

\maketitle 

\vspace{0.5cm}
\begin{abstract}
Prethermalization refers to the transient phenomenon where a system thermalizes according to a Hamiltonian that is not the generator of its evolution. 
We provide here a rigorous framework for quantum spin systems where prethermalization is exhibited for very long times.
First, we consider quantum spin systems under periodic driving at high frequency $\nu$. 
We prove that up to a quasi-exponential time $\tau_* \sim \e^{c \frac{\nu}{\log^3 \nu}}$, the system barely absorbs energy.  
Instead, there is an effective local Hamiltonian $\widehat D$ that governs the time evolution up to $\tau_*$, and hence this effective Hamiltonian is a conserved quantity up to $\tau_*$.
Next, we consider systems without driving, but with a separation of energy scales in the Hamiltonian.  A prime example is the Fermi-Hubbard model where the interaction $U$ is much larger than the hopping $J$.  
Also here we prove the emergence of an effective conserved quantity, different from the Hamiltonian, up to a time $\tau_*$ that is (almost) exponential in $U/J$.


\end{abstract}

\newpage
\section{Introduction}

\subsection{Periodically driven systems}
Time-dependent periodic perturbations arise naturally in various physical systems, e.g. when a physical system is irradiated with electromagnetic fields. 
Such periodically driven quantum systems exhibit rich and often unexpected behaviour~\cite{bukov2015universal}. One classic example is the dynamical localization of a  kicked quantum rotor~\cite{fishman1982chaos}.  
Another well-known (but non-quantum) example is the Kapitza pendulum {\cite{kapitza1951dynamic}}, where a sufficiently fast periodic drive stabilizes the otherwise unstable fixed point in which the pendulum stands on its head. 

More recently, it has been suggested that periodic driving can be used as a tool to design interesting and exotic many-body Hamiltonians. 
The underlying idea here is that the time evolution operator $U(t)$, generated by a periodically modulated Hamiltonian $H(t)= D+V(t)$ with period $T$, could be recast as
\beq   \label{eq: equality}
   U(mT) = \e^{-\i m T  \widehat D  } ~~~\text{ for } m \in \mathbb{Z},
\eeq
with $\widehat D$ the `effective' Hamiltonian, and $U(T)$ sometimes called `monodromy' or Floquet' operator.
A prime example of \eqref{eq: equality} is where the system consists of non-interacting fermions on the lattice: 
$$
H(t)=  \sum_{x,y \in \Lambda}     h(x,y,t) c^*_x c^{}_y,
$$
with $c^{}_x,c^*_x$, fermionic field operators and $h(\cdot,\cdot, t)$ the kernel of a self-adjoint operator on $l^2(\Lambda)$ with the volume $\Lambda$ a finite subset of $\bbZ^d$.  
 We write the one-particle unitary evolution $u(t)$, solving $\frac{\d }{\d t}u(t)=-\i h(t) u(t)$ with $ u(0)=1$.  Then an algebraic exercise yields \eqref{eq: equality} with 
$$
 \widehat D =   \sum_{x,y}   \widehat d(x,y) c^*_x   c^{}_y, \qquad \text{with $\widehat d$ solving $\e^{-\i T \widehat d}= u(T)$}
$$
If $T\norm h(t) \norm \ll 1$, then the spectrum of $u(T)$ covers only a small patch of the unit circle and one can construct $\widehat{d}$ as a convergent series in $h(t),0<t\leq T$. Its leading term (as $T\to 0$) is the averaged Hamiltonian $\frac{1}{T}\int_0^T \d t h(t)$. 
Hence, under these conditions $\widehat D $ can be chosen as a local many-body Hamiltonian that is moreover locally close to  $\frac{1}{T}\int_0^T \d t H(t)$. When calling extensive operators `local', we mean that they can be written as a sum of local terms, the range of which does not grow with volume.  The expansion alluded to is in general known as the 'Magnus' expansion', see e.g.\ \cite{magnus1954exponential,bukov2015universal}. 

Recent theoretical works~\cite{ponte2015periodically,d2014long,lazarides2014equilibrium} suggest, based on the quantum version of the ergodic hypothesis (Eigenstate Thermalization Hypothesis),  that generic periodically driven many-body systems eventually do heat up to an infinite temperature, and so one would {\emph{not}} expect the strict equality \eqref{eq: equality} to hold for a local $\widehat D$. 
Apart from non-interacting systems,  \eqref{eq: equality} can still be true if $D$ describes a disordered many-body localized system, and $T\norm V(t)\norm$ is sufficiently small.  In that case, numerics and theory \cite{ponte2015many,lazarides2015fate,abanin2016theory}   suggest that \eqref{eq: equality} holds with $ \widehat D$ similar to $D$.  
But, as said, this is not our prime interest here and we study systems for which \eqref{eq: equality} is not expected to hold strictly.

In this paper, we argue instead that in great generality, the equality  \eqref{eq: equality} holds however, approximatively, up to the quasi-exponential time $\tau_* \sim \e^{c \frac{\nu}{\log^3 \nu}}$, i.e.\@ for $mT \le \tau_*$, with $\nu=2\pi/T$ the frequency, for a local Hamiltonian $\widehat{D}$.
In particular, we prove that $\widehat{D}$ is approximatively conserved up to time $\tau_*$, 
and that the evolution of local observables $O(t)=U(t)OU^*(t)$  for stroboscopic times $t \in T \bbN$ is well-approximated by the effective Heisenberg evolution $\e^{\i t \widehat{D} } O \e^{-\i t \widehat{D} } $. 

The physical significance of our work is hence that we rigorously underpin the use of effective Hamiltonians $\widehat D$.  Part of the appeal of this idea is that $\widehat D$ can have different properties from the original static Hamiltonian $D$, for example, nonzero Chern numbers, see Ref.~\cite{bukov2015universal,aidelsburger2015measuring,kitagawa2010topological,lindner2011floquet,goldman2014periodically}.

\subsection{Heuristics and connection to non-driven systems}
The emergence of an effective conserved quantity is not specific to periodically driven systems, and we now outline how to generalize our results to closed (i.e.\@ non-driven) Hamiltonian dynamics.  At the same time, we develop the basic intuition that underlies our results. 

\subsubsection{Driven systems in Floquet representation}
To analyze the evolution equation $\i \partial_t \phi (t)  = H(t)\phi (t) $, with $H(t) = H(t+T)$ acting on a Hilbert space $\caH$, a standard technique is to work in an extended space, see \cite{howland1974stationary} for details.
If a sufficiently regular $ \psi (\theta,t) \in  L^2(\mathbb R/ T \mathbb Z, \caH)$ solves the equation
\begin{equation}\label{eq: extended}
\i \partial_t \psi (\theta,t) =  \big( - \i \partial_\theta + H(\theta) \big)  \psi (\theta , t),
\end{equation}
(with $\i \partial_\theta$ defined with periodic boundary conditions) 
then $ \phi(t) = \psi (t,t) $ solves the equation  $\i \partial_t \phi (t)  = H(t) \phi (t)$. Under  Fourier transform $L^2(\mathbb R/ T, \caH ) \mapsto l^2(\bbZ,\caH): \psi(\theta) \to \tilde \psi(k)$, the equation \eqref{eq: extended} reads
\begin{equation} \label{eq: extended eq}
\i \partial_t  \tilde \psi (k,t)  =  k\nu \tilde\psi (k , t)   + \widetilde H \tilde\psi(k,t)
\end{equation}
where $\nu=2\pi/T$ and $\widetilde H$ acts by  convolution with $\widetilde H(k)$.  
The generator of \eqref{eq: extended eq}, acting on the extended space, is given by 
\begin{equation}\label{eq: G Hamiltonian}
G :=  \nu N + D + V . 
\end{equation} 
where $N$ is multiplication by $k$,  
$D$ is the part of $\widetilde H$ that acts diagonally in the $k$-coordinate, by multiplication with
$$
\widetilde H(0) = \int^T_0 \d t H(t)
$$
and $V$ is the remaining part, given by
$$
V\psi(k) =\sum_{k' \neq k} \widetilde H(k-k') \psi(k').
$$

If $\norm V \norm, \norm D \norm \ll \nu$, then spectral perturbation theory would apply and we could construct spectral subspaces for $G$, corresponding to different $k$, as those are separated by large gaps $\nu$.  
In particular, this would lead to the existence of a $\widehat D$ satisfying \eqref{eq: equality}. 
In a generic many-body system, $D,V$ have norm proportional to the volume and spectral perturbation theory should not be expected to apply.  {Before continuing this line of thought, we first present the time-independent setup (closed systems), since there the starting point is similar to the setup we arrived at in \eqref{eq: G Hamiltonian}}

\subsubsection{Closed systems}

The strategy for closed systems is to find a static Hamiltonian that has the same structure and phenomenology as the extended operator $G$ defined above.  Its important features are that there is a term $\nu N$ with large spectral gaps. However, this is not so easy to achieve in a many body system. If we take $N=\sum_x N_x$ a sum of strictly local and commuting terms, then it is not sufficient that $\nu N_x$
 have large spectral gaps, compared to the strength of other local terms.  An obvious possibility is however when the spectrum of all of the $\nu N_x$ is a subset of $\nu \bbZ$ and we will exploit this possibility.
The Fermi-Hubbard model is a natural example
$$
G=     J \sum_{x\sim y, \sigma} c^{}_{x,\sigma} c^*_{y,\sigma}   +   U  \sum_{x}  n_{x, \downarrow} n_{x, \uparrow}=:  H + U N,
$$
with interaction $U$ much larger than $J$, and $U$ playing the role of $\nu$ above. To allow for a unified discussion, we will hence simply write $\nu$ instead of $U$.
Note that the spectrum of  $N_x=n_{x, \downarrow} n_{x, \uparrow} $ here is $1$ (when there are two fermions at $x$, this is a 'doublon') and $0$ (otherwise).  
Our theory expresses rigorously that doublons and singlons (sites with a single fermion) do not interact up to quasi-exponentially long times in $\nu/J$, see also
 \cite{sensarma2010lifetime,chudnovskiy2012doublon,bukov2015prethermal}.
We refer to Section \ref{sec: examples} for more  discussion.

\subsubsection{Non-convergent perturbation theory}

In both of the above setups, the point is that the Hamiltonian $G=\nu N+H$ is dominated by $\nu N$, where $N$ is a sum of local terms, and it has spectral gaps that remain open in the thermodynamic limit. 
As already remarked, it is certainly not true that spectral perturbation theory applies as such, as the norm of the perturbation $H$ grows with volume whereas the gaps do not.  
However, what is true is that matrix elements of the perturbation are smaller than the gaps of $\nu N$. To substantiate this,  let $\str\eta \rangle=\otimes_x \str \eta_x\rangle$ be the product basis of eigenvectors of $N$. Here $\str \eta_x \rangle$ are a basis of the single-site space $\bbC^4$ in which $N_x$ is diagonal, and we write  $N(\eta) \equiv\langle \eta \str N \str \eta\rangle$. Then, the point is that, whenever $N(\eta) \neq N(\eta')$, then  
\beq \label{eq: bounded local perturbation}
\str\langle \eta'\str H \str \eta \rangle \str \leq O(1), \qquad \text{i.e.\ not growing with volume.}
\eeq
This means that 
\beq  \label{eq: non resonance condition}
\frac{\langle \eta'\str H \str \eta \rangle}{ \str \nu N(\eta')-\nu N(\eta) \str}  =\mathcal{O}(1/\nu)  \ll 1
\eeq
That is, perturbation theory seems applicable in first order, and the same analysis can be repeated at higher orders.  Even though this is really a weaker property than convergent perturbation theory, it does have a physical consequence, namely that the transitions caused by the perturbation take place at a rate slower than $1/\nu$. As far as we know, the easiest way to see this is by performing a unitary transformation that eliminates the perturbation in first order, e.g.\ to perform a KAM step, see \cite{imbrie2016many} for a celebrated recent application and an account of the technique.  In our case, we can proceed to eliminate further orders and it is clear that the rate of transitions is actually smaller than any power of $1/\nu$.  The reason why we cannot proceed \emph{ad infinitum} (and indeed, do not believe it would yield a physically correct result) is that in higher orders the operator $H$ in \eqref{eq: bounded local perturbation} gets replaced with an operator whose local terms grow in range and norm as the order increases.  We basically show that only at $n_*$'th order, with $n_*=\nu$, the condition \eqref{eq: non resonance condition} gets violated. Physically, this means that at that order the perturbation connects different subspaces of $N$ resonantly and hence the rate of transition is indeed roughly given by $(1/\nu)^{n_*}$, the strength of the perturbation at that order.
  For a rather intuitive calculation we refer to \cite{abanin2015exponentially}, where we prove that the linear response heating rate is bounded by $\e^{-c\nu}$. The present paper is not restricted to linear response and the final bound on the 'heating rate' is rather  $\e^{-c \frac{\nu}{\log^3 \nu}}$ due to combinatorial factors. However, in dimension $1$, we can obtain  $\e^{-c\nu}$, see \cite{PhysRevB.95.014112}, at the cost of restricting $H$ to be a sum of strictly local terms. 

\subsection{Previous results}

Among our models, the most tractable is certainly the case of non-interacting fermions (strictly speaking, fermionic lattice systems are not covered here, but this could easily be remedied). It is therefore surprising that the phenomenon of 'localization in energy' has barely been rigorously studied in the absence of interaction, i.e.\ for the one-particle case. In \cite{combescure1990spectral}, localization was proved for periodically kicked operators.
The authors of \cite{soffer2003anderson} considered the disordered Anderson model with a local time-periodic perturbation and proved the stability of localization. In \cite{bourgain2004anderson}, the same is achieved for a quasi-periodic perturbation, which can be viewed as the case of multiple frequency dimensions. Very recently, \cite{ducatez2016anderson} considered the problem from a very similar point of view as in the present paper:  \cite{ducatez2016anderson} proves stability of Anderson localization with a periodic driving term, provided that the driving frequency is not too small.
High-frequency asymptotics (instead of strict localization) in periodically driven quantum systems have been investigated in \cite{eckardt2015high,rahav2003effective, verdeny2013accurate} by techniques similar to ours, but not applicable to the many-body problem.

We move now to interacting many-body systems. 
Some recent works proved bounds for linear response heating rate at high frequency, see \cite{bru2017lieb,abanin2015exponentially}.   
 Moreover, when preparing the first version of this manuscript, a result very similar to the present paper appeared: \cite{kuwahara2016floquet,mori2016rigorous} construct an effective Hamiltonian by truncating the Magnus expansion and they prove that it generates the local dynamics up to exponentially long times.  In the case where the driving is localized (as opposed to `a sum of local terms'), \cite{kuwahara2016floquet} contains a much stronger result, namely that the one-cycle unitary of the process is close in operator norm to the unitary generated by an effective Hamiltonian. The main difference between \cite{kuwahara2016floquet,mori2016rigorous} and the present paper, apart from the technique, is that our results are valid for arbitrary dimension.
 
Looking more broadly, we note that there is a large literature on reducibility and quasi-reduciblity of Hamiltonian systems, see e.g.\ \cite{eliasson2009reducibility}. This is in flavour and methodology of course very similar to our work, but the focus is different: reducibility is trivial in the quantum setup as the system is linear, and the only issue is with the locality of the reduced equation, cf. the discussion in the introduction. Finally, our results on closed (time-independent) systems are in a certain sense special cases of a Nekoroshev bound for many-body systems, see \cite{carati2016persistence,de2015asymptotic} and references therein.

\section{Results for driven systems}
\subsection{Setup}
We start from a time-dependent Hermitian Hamiltonian $H(t)=H^*(t)$, acting on $\caH_{\La}= (\bbC^{q})^{\otimes_{\La}}$ with $\La$ a finite subset of the lattice $\bbZ^d$, and $q <\infty$. 

 We are hence dealing with bounded operators throughout, but we will always state bounds that are uniform in the volume $\La$. 
The time-dependence is periodic:
$$
H(t)=H(t+T),
$$
with period $T$ and the mapping $t\mapsto H(t)$ is measurable and bounded.
We split
$$
H(t)=D+V(t),
$$
by setting
$$
D:= \langle H \rangle  :=  (1/T) \int_0^T  H(t), \qquad  V(t) =    H(t)- D.
$$
The dynamics is given by $U(t)$, the unitary family generated by $H(t)$: 
$$
U(t)=-\i \int_0^{t} \d t' H(t') U(t').
$$
which can indeed be solved for measurable, bounded $t\mapsto H(t)$.
Since our assumptions do not guarantee that one can construct an everywhere differentiable $U(t)$, we have not defined $U(t)$ as the solution of $\partial_t U(t)=-\i H(t) U(t)$, but this will not play any role in what follows.

\subsection{Locality and potentials} \label{sec: locality and potentials}
We need some standard notion of locality.  Let $\caB_{\La}$ be the algebra of bounded operators acting on $\caH_{\La}$, equipped with the standard norm $\norm O\norm =\sup_{\psi \in \caH_{\La}, \norm\psi\norm=1 } \norm O \psi \norm$.  We denote by $\caB_{S}\subset \caB_{\La} $, with $S \subset \La$, the subalgebra of operators of the form $1_{\La \setminus S} \otimes O_S$, which is canonically identified with the operators acting on $(\bbC^n)^{\otimes_S}$: we say that $1_{\La \setminus S} \otimes O_S$ `acts within $S$' and in an abuse of notation, also refer to it as $O_S$.
For any operator $Z$ we can decompose (in a nonunique way) $Z= \sum_{S \in 
 \caP_{c}(\La)} Z_S$, where $Z_S \in \mathcal{B}_S$ and $\caP_{c}(\La)$ denotes the set of finite,  connected (by adjacency) subsets of $\La$. The collection  $(Z_S)$ is usually referred to as an (interaction) \emph{potential}, see e.g.\ \cite{nachtergaele2006propagation,simon2014statistical} and in fact, all of our results will be about such potentials. However, to keep the notation simple, we prefer not to make this distinction explicit and to keep the notation $Z$ for operator and potential. In principle, this creates an ambiguity as it is not clear which potential is meant (nonuniqueness of decomposition). In practice, all potentials will be derived in a straightforward way from the potential $(H_S(t))$ of $H(t)$, which is an input to our work.  For example, the above definitions specify $V,D$ by linear operations on $H$, and it is understood that the potentials are given as, for example, $D_S:=(1/T) \int_0^T  H_S(t) $ and $V_S:=H_S-D_S$. 
We will also define below the potential $  \e^{\ad_{A_{n_*}}}\ldots \e^{\ad_{A_0}} Z$ for potentails $A_j,Z$.  To make this explicit, we expand all exponentials in a power series (recall that all our operators are bounded) and we define the 'commutator of potentials $Q,Z$' as
$$
(\ad_{Q}(Z))_{S}:= \sum_{S_1,S_2: S_1 \cup S_2=S}[Q_{S_1},Z_{S_2}]
$$
where we note that only $S_1 \cap S_2 \neq \emptyset$ contribute.  This defines inductively $\e^{\ad_{Q}} Z$. 

We define a family of norms on (time-dependent) potentials, parametrized by a spatial decay rate $\kappa$
$$
\norm Z \norm_{\kappa}  :=    \sup_{x \in \La}  \sum_{S \in \caP_c(\La): S \ni x}  \e^{\kappa \str  S \str}   \sup_{t} \norm Z_S(t) \norm, \quad  \kappa >0
$$
Note that these norms are tailor-made for operators (potentials) that are sums of local terms that themselves are independent of the global volume $\La$, in particular for many-body Hamiltonians.  In principle, one could take $\La$ infinite, i.e.\ $\La=\bbZ^d$. In that case $Z$ still makes sense as a potential (but not as an operator) and $\e^{\ad_{A_j}} $, in case $A_j$ is anti-Hermitian, still makes sense as an automorphism of the $C^*$-algebra of quasilocal operators (but $\e^{A_j}$ does not make sense as a unitary).  
We will not adopt this point of view and we prefer to have a finite $\La$ throughout, so that also the operators remain well-defined. 
\subsection{General results}
 
We will always assume that the frequency $\nu$ is large compared to some local energy scales, namely that there is a decay rate $\kappa_0>0$ such that
\beq \label{ass: initial scales}
\nu \geq \frac{9\pi \norm V \norm_{\kappa_0}}{\kappa_0}, \qquad n_* \geq 1, 
\eeq
where
$$
n_*:=\left\lfloor \frac{\nu/\nu_0}{(1+\ln \nu/\nu_0)^3} \right \rfloor -2, \qquad \text{with} \,\, \nu_0 :=  \frac{54\pi}{\kappa_0^2}  \left( \norm D \norm_{\kappa_0} +  2\norm V \norm_{\kappa_0} \right).
$$
In the theorem below (and further in the text), $C$ refer to numerical constants that can be chosen independent of all model parameters.  Most importantly, they are independent of the volume $\La$. By $K$ we denote numbers that can depend on all model parameter, but not on the frequency $\nu$ and the volume $\Lambda$. 

\bet \label{thm: main}
Assume that \eqref{ass: initial scales} holds,
then there are Hermitian operators (potentials) $\widehat D, \widehat V(t)$ and a unitary $Y(t)$ such that the unitary
$$
\widehat U(t) :=  Y(t) U(t)
$$
solves
$$
 \widehat U(t)  =-\i \int_0^{t}\d t' (\widehat D+ \widehat V(t')) \widehat U(t'),
$$
and the following are satisfied:
\begin{enumerate}
\item  $\widehat D$ is time-independent and $Y(t),\widehat V(t)$ are $T$-periodic. $Y(t)=1$ for stroboscopic times $t \in T  \bbN$.
\item   Set the decay rate $ \kappa_{n_*}:=  \kappa_0 (1+\log (n_*+1))^{-1} $, then 
\beq 
\norm \widehat D - D  \norm_{  \kappa_{n_*}} \leq   C (\nu_0/\nu), \qquad    \norm  \widehat V \norm_{  \kappa_{n_*}} \leq  C(2/3)^{n_*}\norm V  \norm_{  \kappa_{0}}    \label{eq: diff d hatd}. 
\eeq
\item The unitaries $Y(t)$  are defined by  $Y(t)=  \e^{{A_{n_*}(t)}}\ldots \e^{{A_0(t)}}$ with operators $A_j(t)$ to be specified later. They are close to identity
 and quasilocal in the sense that 
 $$ \norm Y(t)ZY^{*}(t)  - Z  \norm_{  \kappa_{n_*}} \leq  C  (\nu_0/\nu)   \norm Z \norm_{\kappa_0}, \qquad \text{ for any operator (potential) $Z$, for all $t\in \mathbb R$.} $$
\end{enumerate}
(we write $Y(t)ZY^{*}(t)= \e^{\ad_{A_{n_*}}}\ldots \e^{\ad_{A_0}} Z$ to interpret it as a  potential, see Section \ref{sec: locality and potentials})
\eet
 For a $Z$ that is local, and not merely a sum of local terms, the last bound of the theorem above is obviously useless; in this case, it actually follows easily from our construction that $Y(t)ZY^*(t)-Z$ has small operator norm. 
Note that the times $t \in T \bbZ$ are, artificially, singled out to play a distinguished role, namely $Y(T\bbZ)=1$.   
A more intuitive way to phrase this result is by going to a periodically rotating frame, given by the unitary $Y(t)$. If $\psi(t)$ solves $\partial_t \psi(t)=-\i H(t) \psi(t)$, then
$\widehat\psi(t)=Y(t)\psi(t)$ solves  $\widehat\psi(t)=-\i\widehat H(t)\widehat\psi(t)$ with 
$$
\widehat H(t)= Y(t)  H(t) Y^*(t) -\i  Y(t) \partial_t Y^*(t).
$$
 Theorem \ref{thm: main} asserts then that the Hamiltonian $\widehat H(t)$ generating this evolution is almost time-independent for large $\nu$, since the time-dependent part $\widehat V$ has local terms of order $(2/3)^{\frac{\nu/\nu_0}{(1+\ln(\nu/\nu_0))^3} }$. 

\subsection{Physical consequences}

As  explained above, our technical result suggests that the evolution is well-described by an effective Hamiltonian, at least for stroboscopic times.
This implies that the effective Hamiltonian $\widehat D$ is a quasi-conserved quantity, see \eqref{eq: slowheating one} in the upcoming theorem.   Since $D$ is close to $\widehat D$, this also implies that $D$ itself is well-conserved, \eqref{eq: slowheating two}.  In other words, apart from a quantity of order $1/\nu$, the energy density grows very slowly. 
\begin{theorem}[Slow heating]\label{prop: slow heating}
\begin{align}
&\frac{1}{\str \La \str} \norm U^*(t) \hat D  U(t) -    \hat D \norm  \leq   t \, K_0\,    (2/3)^{n_*} \qquad \text{for $t \in T \bbN $}, \label{eq: slowheating one} \\[1mm]
&\frac{1}{\str \La \str} \norm U^*(t) D  U(t) -    D \norm  \leq   t \, K_0\,    (2/3)^{n_*}  +  C(\nu_0/\nu),  \qquad \text{for any $t \geq 0 $}  \label{eq: slowheating two}
\end{align}
with $K_0 = C \norm D \norm_{\kappa_0} \norm V \norm_{\kappa_0}  $ and $n_*$ as in Theorem \ref{thm: main}.
\end{theorem}
Note that in this theorem, we use the standard operator norm $\norm \cdot\norm$, but we divide by volume. 
Next, we state how the evolution of local operators is approximated by the evolution with the time-independent Hamiltonian $\widehat{D}$. For a local observable $O$, the difference 
$$
\widehat U^*(t)O  \widehat U(t)  -      \e^{\i t \widehat D}  O    \e^{-\i t \widehat D} 
$$
grows very slowly as it is due to the term $\widehat V$. However, to make this precise, we need to control spatial spreading of $\e^{\i t \widehat D}  O    \e^{-\i t \widehat D}$, which  is done by invoking a  Lieb-Robinson bound.  
   Since $\widehat U(t)=U(t)$ for stroboscopic times, we get then
\begin{theorem}[Approximation of local observables]\label{prop: local approx}
For any rate $0<r<\ln(3/2)$,
there are numbers $K(O),K'(O) <\infty$, depending on model parameters and the observable $O$, but not $\nu$, such that
$$
\norm U^*(t) OU(t) - \e^{\i t \widehat D^{}} O \e^{-\i t \widehat D^{}}   \norm \leq  K(O)   \e^{-r n_{*}} (t+K'(O))^{d+1}, \qquad \text{for $t \in T \bbN$}.
$$
\end{theorem}
This theorem is for us the most clear expression of the fact that $\widehat D$ really describes the dynamics for very long times.
We do not see how to improve the dependence of the bound on $t$, unless one would manage to replace the Lieb-Robinson bound by a diffusive bound. Phrased in a different way, this result says that the dynamics generated by $\widehat D$ is close to the actual dynamics up to a time that grows quasi-exponentially in $\nu$. 
Theorem \ref{prop: local approx} is actually a particular case of a more general statement valid for any $t\in \mathbb R$, provided that
$ \e^{\i t \widehat D^{}} O \e^{-\i t \widehat D^{}}  $ {is replaced} by  $\e^{\i t \widehat D^{}} Y(t)OY^*(t) \e^{-\i t \widehat D^{}} $.  

\section{Results for time-independent systems}

\subsection{Setup}

We turn to a time-independent setup.

Let us have a family of `number operators' $N_x$ acting on site $x$. By number operators, we simply mean that $\sigma(N_x)\subset \bbZ$, for every $x$, with $\sigma(\cdot)$ the spectrum.  Set 
$$
N:= \sum_x N_x
$$
The idea is that a multiple of this operator dominates the Hamiltonian of our system; the total Hamiltonian is
$$
G:= \nu N+H
$$
with $\nu$ large compared to the local energy scales of $H$.  We decompose $H=D+V$ with 
$$
D:=\langle H \rangle = \sum_{n \in \bbZ}  \chi(N=n) H  \chi(N=n) , \qquad   V = H-\langle H \rangle
$$
such that $[N,D]=0$. 
The choice of not simply calling the total Hamiltonian $H$ is to make the analogy with the time-dependent case maximal, as will be clear later.
Relevant examples of this setup will be discussed below. 

\subsection{General result}
To state the result, we now exploit the notational similarity with the time-dependent case, which allows for most definitions and formulas to be identical in both cases. The norms $\norm \cdot \norm_{\kappa}$ are defined as before, except that the supremum over time is now omitted (there is no time-dependence).  

We assume that the parameter $\nu$ is large compared to some local energy scales, namely that there is a decay rate $\kappa_0>0$ such that
\beq \label{ass: initial scales closed}
\nu \geq \frac{9\pi \norm V \norm_{\kappa_0}}{\kappa_0}, \qquad n_* \geq 1, 
\eeq
where
$$
n_*:=\left\lfloor \frac{\nu/\nu_0}{(1+\ln \nu/\nu_0)^3} \right \rfloor -2, \qquad \text{with} \,\, \nu_0 :=  \frac{54\pi}{\kappa_0^2}  \left( \norm D \norm_{\kappa_0} +  2\norm V \norm_{\kappa_0} \right).
$$
In the theorem below (and further in the text), $C$ refer to numerical constants that can be chosen independent of all model parameters. Most importantly, they are independent of the volume $\La$.

\bet \label{thm: mainstatic}
Assume that \eqref{ass: initial scales closed} holds,
then there are Hermitian operators (potentials) $ \widehat H, \widehat D, \widehat V$ and a unitary $Y$, such that
$$
Y (\nu N+ H) Y^*= \nu N+ \widehat H =  \nu N+ \widehat D+\widehat V 
$$
with 
\begin{enumerate}
\item  $\widehat D= \langle \widehat H \rangle$, i.e.\ $[\widehat D, N]=0$. 
\item   Set the decay rate $ \kappa_{n_*}:=  \kappa_0 (1+\log (n_*+1))^{-1} $, then   
\beq 
\norm \widehat D - D  \norm_{  \kappa_{n_*}} \leq   C (\nu_0/\nu), \qquad    \norm  \widehat V \norm_{  \kappa_{n_*}} \leq  (2/3)^{n_*}\norm V  \norm_{  \kappa_{0}}    \label{eq: diff d hatd static}. 
\eeq
\item The unitaries $Y$ are close to the identity and quasilocal in the sense that, for any operator (potential) $Z$,
 $$ \norm YZY^{*}  - Z  \norm_{  \kappa_{n_*}} \leq C  (\nu_0/\nu)   \norm Z \norm_{\kappa_0}.$$
\end{enumerate}
\eet

\subsection{Physical consequences}
Thanks to the above theorem, we identify two extensive quantities that stay almost conserved for quasi-exponentially long times in $\nu$. 
Let 
$$\mathcal{D} = Y^* \widehat D Y, \qquad \mathcal N = Y^* N Y .$$
The following result derives from Theorem \ref{thm: mainstatic} in the same way Theorem \ref{prop: slow heating} follows from Theorem \ref{thm: main}.
\bet \label{thm: conservation d}
For any $t\geq 0$,
\begin{align}
& \frac{1}{\str \La \str} \norm U^*(t) \mathcal D U(t)  - \mathcal D \norm  \leq  t K_0 (2/3)^{n_*},  \label{eq: slowheatingstatic} \\
&  \frac{1}{\str \La \str} \norm U^*(t) \mathcal N U(t)  - \mathcal N \norm  \leq  t K'_0 (2/3)^{n_*},   \label{eq: slowheating2static}
\end{align}
with $K_0 = C \norm D \norm_{\kappa_0} \norm V \norm_{\kappa_0}  $, $K'_0 = C (\sup_x\norm N_x \norm)  \norm V \norm_{\kappa_0} $,  and $n_*$ as in Theorem \ref{thm: mainstatic}. 
\eet
Let us notice that 
\beq  \label{eq: n dominates h}
N =  \frac{1}{\nu}G -  \frac{1}{\nu}H
\eeq
so $N$ simply equals a conserved quantity plus a term of norm {proportional to }$\str \La \str/\nu$; hence $N$ is always conserved up to an error of order $|\Lambda|/\nu$. 
Our second theorem states that the evolution of local observables is well-described by the effective Hamiltonian
$ \nu N + \widehat D$, modulo an error of order $1/\nu$ and up to quasi-exponentially large times. 
\bet  For any $0<r_1<\tfrac{1}{d+1} \ln(3/2)$,  and local operator $O$, there is a $K_3(O)$ such that
$$
\norm U^*(t)  O U(t)  -  \e^{\i t (\nu N + \widehat D)}  O \e^{-\i t (\nu N + \widehat D)}  \norm  \leq  \frac{1}{\nu} K_3(O), \qquad \text{for $t \leq \e^{r_1 n_*}$}. 
$$
\eet
To prove this theorem, we start from
$$
Y U^*(t) Y^* OY U(t)  Y^* =  \e^{\i t (\nu N + \widehat H)}  O \e^{-\i t (\nu N + \widehat H)} 
$$
en then use Lieb-Robinson bounds to replace the right hand side by  $\e^{\i t (\nu N + \widehat D)}  O \e^{-\i t (\nu N + \widehat D)}  $, up to an error that remains small for $t \leq \e^{r_1 n_*}$. This is analogous to the proof of  Theorem \ref{prop: local approx}. Then, one gets rid of the $Y,Y^*$ by using that these unitaries are locally close to identity, in particular, 
$ \norm Y^* OY -O \norm \leq  \frac{1}{\nu}K(O) $. Such statements are similar to  Statement 3) of Theorem \ref{thm: mainstatic} and their  proof is an obvious variation.

\subsection{Examples}\label{sec: examples}

\subsubsection{Fermi-Hubbard chain}
We consider the Fermi-Hubbard Hamiltonian, for $J, U \in \bbR$, 
$$
G=     J \sum_{x\sim y, \sigma} c^{}_{x,\sigma} c^*_{y,\sigma}   +   U  \sum_{x}  n_{x, \downarrow} n_{x, \uparrow} 
 $$
where  $\sigma\in \{\uparrow, \downarrow\}$, $c^{}_{x,\sigma}, c^*_{x,\sigma}$ are fermionic field operators, i.e.\ 
$$
\{c^{}_{x,\sigma}, c^*_{x',\sigma'} \} =  \delta_{x,x'}\delta_{\sigma, \sigma'}, \qquad  \{c^{}_{x,\sigma}, c^{}_{x',\sigma'} \} =0, \qquad  \{c^*_{x,\sigma}, c^*_{x',\sigma'} \} =0
$$
and $n_{x,\sigma}=c^*_{x,\sigma}c^{}_{x,\sigma} $ and $n_x= \sum_{\sigma}n_{x,\sigma}$.
Apart from the energy, there are two conserved quantities, namely $n_\sigma=\sum_{x} n_{x,\sigma}$. 
We now assume that $U \gg J$.    We set $\nu \equiv U$ and 
$$
N_x :=  n_{x, \downarrow} n_{x, \uparrow}  =  \chi(n_x =2),  \qquad   N=\sum_x N_x,
$$ 
i.e.\ $N$ is the number of doublons in the system.   
Now we find $D$ as the part of $H$ that commutes with $N$: 
$$
D=  T_{\text{s}} + T_{\text{d}} :=  J \sum_{x\sim y, \sigma} c^{}_{x,\sigma} c^*_{y,\sigma}  \, \chi(n_x+n_{y}=1)  +  J \sum_{x\sim y, \sigma} c^{}_{x,\sigma} c^*_{y,\sigma}  \, \chi(n_x+n_{y}=3) 
$$
where $T_{\text{s}}, T_{\text{d}}$ stand for the `singlon' and `doublon' kinetic energies, respectively.  In fact, both $T_{\text{s}}, T_{\text{d}}$ commute with $N$. 
The total Hamiltonian is hence 
$
\nu N+  T_{\text{s}} + T_{\text{d}} + V,
$
in accordance with the general abstract setup. 

As observed in \eqref{eq: n dominates h}, the density of doublons, $\frac{N}{\str \La\str}$, is conserved up to  a quantity of order $\caO(1/\nu)$ by energy conservation. 
According to Theorem \ref{thm: conservation d}, we conclude that $\mathcal N$, a dressed version of the number of doublons $N$, as well as $ D =  T_s +  T_d$ 
are extensive quasi-conserved quantities, up to an (almost) exponentially small quantity for (almost) exponentially long times in $\nu$.   
This remarkable feature shows thus the appearance of a long pre-thermal regime in the Fermi-Hubbard chain in the regime where $J/U \ll 1$.

\subsubsection{XYZ chain with large magnetic field}
We consider the spin $s$-chain with Hamiltonian
$$
G=  \sum_{x}  J_1S^1_xS^1_{x+1} +  J_2S^2_xS^2_{x+1}+J_3S^3_xS^3_{x+1} +  h\sum_x S^{3}_x
$$
with $S^{\alpha}$  the spin-$s$ representation of $SU(2)$ acting on $\bbC^{2s+1}$ and $S^{\alpha}_x$ copies thereof on site $x$. 
We choose $h \gg J_{\alpha}$ and we set $\nu\equiv h$ and 
$$
N_x :=  S^{3}_x, \qquad \text{i.e.\ the magnetization}
$$
The operator $D$ is then given by 
$$
D= \sum_{x}  J(S^1_xS^1_{x+1} + S^2_xS^2_{x+1}) + J_3S^3_xS^3_{x+1}, \qquad  2J= J_1+J_2
$$
which indeed commutes with the magnetization $N$.
So we see here that there is an emergent $U(1)$ symmetry in $D$, corresponding to the conservation of $N$.


%
%
%
%
%

%
%

\section{Renormalization of Hamiltonians} \label{sec: ren ham}

\subsection{Recursion formulae}

We now describe our main scheme to transform the Hamiltonian. It is inspired by \cite{bambusi2001time,imbrie2016many,abanin2016theory}.
Let us rename the operators $H(t),D,V(t)$ as  $H_0(t),D_0,V_0(t)$ and we will now construct $H_n(t),D_n,V_n(t)$, with $n$ up to $n_*$. 
We mostly drop the $t$ in the notation and it is simply understood that  $D_n$ are time-independent, whereas other operators are $T$-periodic. 
At each scale we define $D_n,V_n$ from $H_n$ by setting
$$
D_n:= \langle H_n \rangle  :=  (1/T) \int_0^T  H_n(t), \qquad  V_n=    H_n- D_n.
$$
 $H_{n+1}$ is constructed out of $H_n$ by
\begin{align} \label{def: hk}
H_{n+1} &:= \e^{- A_n}  H_n   \e^{ A_n}  - \i\e^{- A_n}  \partial_t    \e^{ A_n}
\end{align}
where  $A_n$ is determined by
\beq \label{def: ak}
V_n -\i \partial_t A_n=0,  \qquad  A_n(t=0)=0
\eeq
and we note that $A_n$ is indeed $T$-periodic because $\langle V_n \rangle=0$. Note that though we have not demanded $t \mapsto V(t)$ to be smooth, but simply bounded and measurable, the use of derivatives above is justified. Indeed, the integral of a bounded, measurable function $f$ is differentiable  almost surely (because it has bounded variation on intervals) and the derivative equals $f$ almost surely.

We have now defined all operators $H_n$. To appreciate why such a procedure is useful, i.e.\ why the $V_n$ decrease with $n$, we unwrap the recursion relation a bit. 
We define the transformations ($O$ is an arbitrary operator)
$$
  \ga_n (O) :=  \e^{- A_n}   O     \e^{ A_n}  =    \e^{- \mathrm{ad}_{A_n}}   O. 
$$
and 
  $$
  \alpha_n(O)  :=   \int_0^1 \d s \,   \e^{- s A_n}   O     \e^{ s A_n}  =    \int_0^1 \d s \,   \e^{-s \mathrm{ad}_{A_n}}   O. 
  $$
 The latter involves a dummy time $s$ that has nothing to do with the cycle time $t$; the transformation $\alpha_k$ is defined pointwisely for any $t $ in the cycle. 
The use of $\alpha_n$ is that
$$
\e^{- A_n} \partial_t  \e^{ A_n} =   \alpha_n(\partial_t A_n)
$$
as one easily checks by an explicit calculation. If $A_n(t)$ for different $t$ would commute among themselves, then we would simply find back the familiar expression
$$
\e^{- A_n} \partial_t  \e^{ A_n} =   \partial_t A_n, \qquad  \text{(wrong in general})
$$
Recasting  \eqref{def: hk} with the help of the above notation, we get
\begin{align}
H_{n+1} & =  \gamma_n(H_n) -  \i \alpha_n(\partial_t A_n) \\[2mm]
&=      \gamma_n(D_n)  +   ( \gamma_n(V_n)-V_n)  +  (V_n - \i \partial_t A_n)    - \i  ( \alpha_n(\partial_t A_n)  -\partial_t A_n)   \\[2mm]
& =      \gamma_n(D_n)  +   ( \gamma_n(V_n)-V_n)  -   ( \alpha_n(V_n)  -V_n).
\end{align}
For later convenience, we remark that, upon splitting $H_{n+1}=D_{n+1}+V_{n+1}$,
\begin{align} \label{eq: dep on w}
D_{n+1}=D_n +\langle W_{n} \rangle, \qquad  V_{n+1} & =   W_{n}- \langle W_{n} \rangle,
\end{align}
with 
\begin{align} \label{def: wk}
  W_n :=    (\gamma_n(D_n) -D_n)  +   ( \gamma_n(V_n)-V_n)  - ( \alpha_n(V_n)  -V_n).
\end{align}
This defines the iteration scheme and we now move to bounds.

\subsection{Iterative bounds}

The  relation  \eqref{def: wk} makes it particularly intuitive that $W_{n}$ and hence $V_{n+1}$ are of higher order in $\su$ than $V_n$.  
Indeed, from \eqref{def: ak} and periodicity of $V_n(t)$,
$$
\norm A_n \norm_{\kappa}  \leq  (T/2)  \norm V_n \norm_{\kappa}, \qquad \text{for any $\kappa>0$}
$$
 and \eqref{def: wk} contains three exponentials of $A_k$ with zero'th order terms removed, so that keeping only the first order suggests a bound like
\beq  \label{eq: naive iteration bound}
\norm V_{n+1} \norm_{\kappa}  \approx (T/2) \norm V_n \norm_{\kappa} \left(   \norm D_n \norm_{\kappa}+ 2\norm V_n \norm_{\kappa}\right).
\eeq
To make this precise, we will allow the decay rate $\kappa$ on the left-hand side to be slightly smaller than that on the right hand side.
We consider a family of norms by fixing a strictly decreasing sequence of decay rates $\kappa_n>0, n\geq 1$. 
Eventually we will choose $\kappa_n=(1+\log (n+1))^{-1}\kappa_0$ with $\kappa_0$ chosen appropriately for the initial Hamiltonians (cfr.\ statement of Theorem \ref{thm: main}), but for the time being it is convenient to keep the sequence $\kappa_n$ general. We will abbreviate $\norm \cdot \norm_{\kappa(n)}$ by   $\norm \cdot \norm_{n}$.  The following lemma is our prime (and only) tool: 
\begin{lemma}\label{lem: main}
Let $Z, Q$ be potentials and assume that $ 3 \norm Q\norm_{n}    \leq  \delta \kappa_n := {\kappa_{n}}- {\kappa_{n+1}}$. Then 
$$
\norm \e^{Q}Z\e^{-Q}-Z   \norm_{n+1}  \leq      \frac{18}{ \delta \kappa_n \kappa_{n+1}} \norm Z\norm_{n}    \norm Q\norm_{n}.   
$$
Since $ \norm Z\norm_{n+1} \leq  \norm Z\norm_{n}$, we also get
$$
\norm \e^{Q}Z\e^{-Q}   \norm_{n+1}  \leq  \big(1+      \frac{18}{ \delta \kappa_n\kappa_{n+1}}   \norm Q\norm_{n}\big)     \norm Z\norm_{n}.
$$
\end{lemma}
We postpone the combinatorial proof, relying on cluster expansions, to Section \ref{sec: proofs}.

\subsection{Bounds on transformed potentials} \label{sec: bounds on transformed potentials}
We set
$$
v(n):= \norm V_n \norm_n, \qquad  d(n):=    \norm D_n \norm_n, \qquad  \delta d(n):=   \norm D_{n+1}- D_{n}  \norm_{n+1}.
$$
We do not need to introduce any shorthand for $\norm A_n \norm_{n}$ since $\norm A_n \norm_{n} \leq (T/2) v(n) $, as follows from \eqref{def: ak}. 
From \eqref{def: wk} and  Lemma \ref{lem: main},  we then get
$$
\norm W_n \norm_{n+1} \leq  \frac{T}{2} m(n)  v(n) \left[d(n)+2 v(n)\right], \qquad m(n) :=  \frac{18}{\delta\kappa_{n}\kappa_{n+1}}
$$
provided that $(3T/2)  v(n)    \leq \delta{\kappa_{n}}$.
In that case, \eqref{eq: dep on w} yields 
$$
 2\delta d(n), v(n+1)   \leq  T m(n)  v(n) \left[d(n)+2 v(n)\right] 
$$
To get recursive bounds, it is handy to demand that 
\beq \label{eq: condition two thirds} Tm(n)\left[d(n)+2 v(n)\right] <2/3,
\eeq because then   
\beq \label{eq: exp reduction}
v(n+1), \delta d(n) \leq (2/3) v(n)
\eeq
 and hence 
\beq \label{eq: bound dplusv iteration}
d(n+1)+2 v(n+1)  \leq d(n)+2 v(n)
\eeq
which makes it easy to check the validity of \eqref{eq: condition two thirds} at the next order.  Indeed, we see that if 
\beq \label{eq: condition to continue} 
 (3T/2) m(j)  \left[ d(0)+2 v(0)  \right]  \leq  1, \qquad   (3T/2)  v(j)    \leq \delta{\kappa_{j}},  \qquad \text{for any  $0\leq j\leq n$ }
\eeq
then
$$
\delta d(n), v(n+1)  \leq   v(0)  (2/3)^{n+1}.
$$
Let us now use the specific choice  $\kappa_n:=  \frac{\ka_0}{1+ \ln (n+1)}$, then  the second condition of \eqref{eq: condition to continue} is always satisfied provided that 
$$
(3T/2) C_0 v(0) \leq \kappa_0, \qquad \text{with $C_0=\sup_{n\geq 0} (2/3)^n (n+2)\ln(2+n) \leq 3 $}
$$
Thus we find that the conditions
$$
9\pi  v(0)\kappa_0^{-1}  \leq  \nu, \qquad \nu_0 \leq  \nu, \qquad \text{with} \,\, \nu_0 :=  \frac{54\pi}{\kappa_0^2}  \left[ d(0)+2 v(0)  \right] 
$$
allow us to start the procedure and we can continue, i.e.\ \eqref{eq: condition to continue}  remains satisfied, i.e.\ at least as long as $(1+\ln(2+n))^3(n+2) \leq \nu/\nu_0$, hence at least up to  $n=n_*$ with 
$$
n_*=\left\lfloor \frac{\nu/\nu_0}{(1+\ln \nu/\nu_0)^3} \right \rfloor -2.
$$
\subsection{Proof of Theorem \ref{thm: main}}
The above calculation completes an important part of the proof of Theorem \ref{thm: main}, namely the bound on $\norm V_n \norm_n$ at $n=n_*$.  The bound on   ${\widehat D}$ follows by summing the $d(n)$. 
Let us now derive the stated bounds on $Y$. 
By repeated application of Lemma \ref{lem: main} and the bounds derived above, we have the first inequality in (for some $c>0$)
\beq \label{eq: bound multiple transform}
\norm \e^{\ad_{A_{n}}}\ldots \e^{\ad_{A_0}} Z \norm_{n+1} \leq   \norm Z \norm_{0} \prod_{j=0}^{n}(1+ C (\nu_0/\nu)\e^{-cj})  \leq  C \norm Z \norm_{0}
\eeq
The second inequality follows because $(\nu_0/\nu) \leq 1$ (and redefining $C$).  Let
$$
E_{n+1}:= \e^{\ad_{A_{n}}}\ldots \e^{\ad_{A_0}} Z - Z,
$$
so that
$$
E_{n+1}-E_{n}= \e^{\ad_{A_{n}}} (\e^{\ad_{A_{n-1}}}\ldots \e^{\ad_{A_0}} Z )- (\e^{\ad_{A_{n-1}}}\ldots \e^{\ad_{A_0}} Z )
$$
Then we use again Lemma \ref{lem: main} and \eqref{eq: bound multiple transform} (together with the bounds leading to \eqref{eq: bound multiple transform})  to get 
$$
\norm E_{n+1}- E_{n} \norm_{n+1}  \leq  C\e^{-cn} (\nu_0/\nu) \norm Z \norm_{0} 
$$
Item 3) of Theorem \ref{thm: main} follows now by
$$
\norm E_{n_*+1} \norm_{n_*+1} \leq \sum_{j=0}^{n_*} \norm E_{j+1}- E_{j} \norm_{j+1}  \leq  C (\nu_0/\nu) \norm Z \norm_{0}.  
$$

%

\section{Proofs}

\subsection{Proof of Lemma \ref{lem: main}}  \label{sec: proofs}
As in the previous section, we drop the dependence on $t$ from the notation. Starting from \eqref{eq: set p}, one checks that we get indeed the correct bounds on potentials when reinstating (the supremum over) $t$. 
 
We assume that $Q\neq 0,Z \neq 0$, else the claim is trivial. All sets $S \subset \La$ that appear below are assumed to be connected. 
 We expand the exponential
$$
\e^{\ad_Q} (Z_{S_0}) =\sum_{m=0}^{\infty}   \frac{1}{m!}  \sum_{S_1, \ldots S_m}  \,\,   \ad_{Q_{S_m}}   \ldots   \ad_{Q_{S_2} } \ad_{Q_{S_1} } Z_{S_0}.
$$
The integrand vanishes unless $S_j$ has nonempty overlap with $\cup_{i=0}^{j-1} S_i$. 
Localising and taking norms, we get
\beq \label{eq: set p}
\norm (\e^{\ad_Q}Z_{S_0}-Z_{S_0})_{\union} \norm   \leq \sum_{m=1}^{\infty}  \frac{1}{m!}  \sum^{c,\union}_{S_1, \ldots, S_m}  \norm Z_{S_0}\norm \prod_{j} (2\norm  Q_{S_j} \norm),
\eeq
where  $\sum^{c,\union}_{\ldots}$ indicates that the family of sets $S_0,\ldots, S_m$ is connected (they cannot be split into two nonempty mutually disjoint collections) and the union is $\union=\cup_{j=0,\ldots,m}S_j$. 
Multiplying with $\e^{\kappa_{n+1}  \str \union \str}$ and summing over $S_0$, we get 
\beq \label{eq: expansionm}
 \e^{\kappa_{n+1} \str \union \str} \norm (\e^{\ad_Q}Z-Z)_\union \norm    \leq \sum_{m=1}^{\infty}  \frac{\e^{-\kappa(n+1)m}}{m!} (\frac{3}{\delta \kappa_n})^{m+1} \norm Z \norm_{n} \norm Q\norm^m_{n})   \sum^{c,\union}_{S_0, \ldots S_m}  \prod_{j} v(S_j)  
\eeq
where we introduced the shorthand
$$
v(S) :=  (\frac{\delta \kappa_n}{3})\,  {\e^{\kappa(n+1) \str S \str }} \, \left(   \frac{\norm Z_S\norm }{\norm Z\norm_{n}} + \frac{ 2 \norm Q_S\norm }{\norm Q\norm_{n}   } \right)
$$
and we exploited the fact that 
$$
\str \union \str \leq -m+ \sum_{j=0}^{m} \str S_j \str
$$ 
Next,  we use the assumption that $3\norm Q \norm_{n} \leq \delta\kappa_n $ together with $\e^{-\kappa(n+1)m} \leq \frac{2}{\kappa(n+1)(m+1)}$ to get
$$
\text{\eqref{eq: expansionm}} \leq  \underbrace{2 (3/\delta \kappa_n)^2(\kappa(n+1))^{-1} \norm Z \norm_{n} \norm Q\norm_{n})}_{=: L_n} \sum_{m=1}^{\infty}  \frac{1}{(m+1)!}  \sum^{c,\union}_{S_0, \ldots S_m}  \prod_{j} v(S_j).
$$
Taking the sum over $\union$ with $\union \ni x$, we get then 
\beq
 \norm \e^{\ad_Q} Z-Z \norm_{n+1} 
 \leq   L_n  \sup_x \sum_{m=1}^{\infty}   \frac{1} {(m+1)!}  \sum^{c}_{S_0, \ldots S_m} \left(\sum_{j=0}^{m}\chi(x \in S_j)\right)  \prod_{j=0}^m v(S_j).  
\eeq
where '$c$' in the sum $\sum^{c}_{S_0,\ldots, S_m}$ indicates, as above, that the collection of sets is connected.
To perform the sum over connected graphs rooted in $S_j$, we use some standard combinatorial tools: polymer expansions. 
Note that by the definition of the $\norm \cdot \norm_{n}$, we have 
\beq \label{eq: koteckypreiss}
\sum_{S: S\cap S' \neq \emptyset} v(S)  \exp{(\delta \kappa_n \str S \str)} \leq  \delta\kappa_n\str S' \str \qquad \text{for any $S'$},
\eeq
and this assures that one can inductively sum the graphs. A convenient reference is \cite{ueltschi2004cluster}.  Theorem 1 in \cite{ueltschi2004cluster} (more precisely, eq.\ $(5)$ in the proof of said theorem) states that
$$
 \sum_{n=1}^{\infty}   \frac{1} {(m+1)!}  \sum^{c}_{S_0, \ldots S_m} \left(\sum_{j=0}^{m}\chi(x \in S_j)\right)  \prod_{j=0}^m v(S_j) \leq  \delta\kappa_n.
$$
To translate the relevant result there to our work, we identify $\bbA$ with the set of connected subsets of $\Lambda \subset \bbZ^d$, the measure $\mu$ on $\bbA$ with the discrete measure with weights $v(S)$, the function $-\zeta(\cdot,\cdot)$ with the indicator that two sets are non-disjoint and the function $a(S)= \delta \kappa_n\str S\str$. The criterion  $(3)$ in \cite{ueltschi2004cluster} is then precisely \eqref{eq: koteckypreiss} and we use it for singletons $S'=\{x\}$.

\subsection{Proofs of Corollary \ref{prop: slow heating}}

By Theorem \ref{thm: main}, 
$$
U^*(t) Y^*(t) \widehat D Y(t)   U(t) -      \widehat D  =    \i \int_0^t \d s \,    \widehat U^*(s) \,  [\widehat V(s), \widehat D]  \,  \widehat U(s),
$$
and hence the operator norm of the above expression is bounded by
\begin{align}
\int_0^t \d s  \norm [\widehat V(s), \widehat D]  \norm &\leq   2t \str \La \str  \norm \widehat D  \norm_{0} \norm \widehat V \norm_{0}  \\[1mm]
  &\leq   Ct \str \La \str  \norm D \norm_{\kappa_0} \norm V \norm_{\kappa_0}  (2/3)^{n_*},  \label{eq: bound dhat heating}
 \end{align}
 where the first inequality is a straightforward calculation and the second is item 2) of Theorem \ref{thm: main}.  For stroboscopic times $t \in T \bbN$, for which $Y(t)=1$, this gives  \eqref{eq: slowheating one}. 
 To get \eqref{eq: slowheating two}, we use \eqref{eq: bound dhat heating} together with a bound on $\norm D-\hat D \norm$ following from \eqref{eq: diff d hatd}, and a bound on $\norm D- Y(t)DY^*(t) \norm$ following from item 3) of Theorem \ref{thm: main}.

\subsection{Proof of Theorem \ref{prop: local approx} }  \label{sec: proof of corollary}

We use the Duhamel formula to write
\beq   \label{eq: start lr}
\widehat U^*(t) O\widehat U(t) - \e^{\i t \widehat D} O \e^{-\i t \widehat D } =  \int_0^t \d s \,    W^*(s)   [ \widehat V(s)  ,\e^{\i s \widehat D}O \e^{-\i s \widehat D}   ]   W(s),
\eeq
with   $W(s)  =  \widehat U(s)^{-1} \widehat U(t) $.   Hence, to get a bound on the left-hand side, it suffices to estimate the norm of the commutator. The latter is small by the smallness of $\widehat V(s)$, but we also need a Lieb-Robinson bound \cite{lieb1972finite} to avoid getting terms proportional to volume:
\begin{lemma}[Lieb-Robinson bound]  \label{lem: lr}
Let $Z=\sum_S Z_S$ be a  Hermitian  operator (potential) with $ \norm Z\norm_{2\kappa} <\infty $ for some $\kappa>0$.
Let the operators $A,B$ act within $X,Y \subset \La$, respectively. Then,
$$
\norm [A,\e^{\i t Z}B\e^{-\i t Z}] \norm \leq \norm A \norm \norm B \norm \e^{-\kappa (d(X,Y)-v t)} \min(\str X \str, \str Y \str)
$$
with Lieb-Robinson speed $v=v(Z,\kappa):= C(d) (\kappa^{-(d+2)}\e^\kappa) \norm Z\norm_{2\kappa}$ and $C(d)$ only depending on the spatial dimension $d$. 
\end{lemma}
This is a direct consequence of Theorem $2.1$ (in particular, eq.\ 2.17 following it) in \cite{nachtergaele2006propagation}, with obvious adaptations to the notation and setup of the present paper (for example; the function $F$ is chosen as $F(r)=\e^{-\kappa r}(1+r)^{-(d+1)}$). 

We start from \eqref{eq: start lr} and we choose $O$ supported in a set $S_0$. By unitarity of $W(s)$,  we bound the right-hand side as 
\beq  \label{eq: lrfirst}
 \sum_S \norm  [ \widehat V_S(s),   \e^{\i s \widehat D^{(r)}}O\e^{-\i s \widehat D^{(r)}}]  \norm
\eeq
and we can bound the summands by either the above Lieb-Robinson bound, or a trivial norm bound, yielding 
\beq \label{eq: two bounds}   \str S\str \e^{-\kappa (d(S,S_0)-v s)} \, \norm \widehat{V}_S(s) \norm \norm O \norm, \qquad \text{or} \qquad   2 \norm \widehat{V}_S(s) \norm \norm O \norm.
\eeq
with the LR velocity $v=v(\widehat D, \kappa)$, see theorem above. We estimate \eqref{eq: lrfirst} by 
\beq \label{eq: integrals}
C(1 +\tfrac{1}{\kappa}) \str S_0 \str  \norm \widehat{V} \norm_{\kappa} \norm O \norm  \int_0^t \d s \,   \xi^{d}, \qquad   \xi :=  vs+\str S_0\str
\eeq
 using the  first and second bound of \eqref{eq: two bounds} for $S$ such that $d(S,S_0)-v s \leq 0$ and $d(S,S_0)-v s > 0$, respectively. 
%
Now we choose $2\kappa=\kappa_{n_*}$, and the bound on $\norm\widehat{V}\norm_{\kappa_{n_*}}$ from Theorem \ref{thm: main}, to bound \eqref{eq: integrals} by $K(O) \e^{-kn_*} (t+K'(O))^{d+1} $, with  $K(O),K'(O)$ depending on our model parameters and $O$, but not on $\nu$ and $t$. 

 Taking then $t\in T\bbN$ so that $\widehat U(t)=U(t)$, we get the theorem from \eqref{eq: start lr}.

\subsection{Proofs for the time-independent setup}

{Let us see that the algebra in the time-independent setup is identical to the one in the time-dependent problem, so that the proof of the theorems can be taken over in a straightforward way.}
$$
G = \nu N+H=\nu N+D+V, \quad  \text{where} \quad  [D,N]=0
$$
with $\nu$ large compared to the local energy scales of $H$.  The goal is to transform this Hamiltonian into 
$$
\nu N+\widehat H= \nu N +\widehat D+\widehat V
$$ 
with $\widehat V$ quasi-exponentially small in $\nu$ and $\widehat D$ commuting with $N$. 
As in the {time-dependent case}, let us rename $H,D,V$ as $H_0,D_0,V_0$ and we now construct inductively the sequence of operators $H_n,D_n,V_n$. At each step, the splitting $H_n=D_n+V_n$ is defined by 
$$
D_n =\langle H_n \rangle, \qquad V_n=H_n-D_n,
$$
with 
$$\langle O \rangle= (1/T)\int_0^T \d t \, 
\e^{\i t \nu N} O\e^{-\i t \nu N},     \qquad  T:= {2\pi/ \nu}.$$
Indeed, this operation eliminates any off-diagonal elements in $N$-basis, hence we have $[\langle O \rangle,N]=0$. 
The renormalization step defining $H_{n+1}$ is 
$$
\e^{-A_n}(\nu N+H_n) \e^{A_n} = \nu N+H_{n+1}
$$
with $A_n$ determined so that it satisfies the equation
$$
[\nu N,A_n]+V_n=0.
$$
We choose the solution given by
$$
A_n:=  - \frac{\i}{T}  \int_0^T \d t \int_0^t \d s  \e^{\i s \nu N} V_n\e^{-\i s \nu N} 
$$
Indeed, we find that 
$$
[\nu N,A_n]= - \frac{\i}{T}  \int_0^T \d t \int_0^t \d s  \frac{\d }{\d s} (\e^{\i s \nu N} V_n\e^{-\i s \nu N}) =  -\langle V_n\rangle +  V_n  =  V_n
$$
where the last equality $\langle V_n\rangle=0$ is by the definition of $V_n$. 
A straightforward bound gives 
$$
\norm A_n \norm \leq   \frac{T}{2}  \norm V_n\norm 
$$

Defining now $\gamma_n, \alpha_n$ just as before, i.e.\  $$
  \ga_n (O) :=  \e^{- A_n}   O     \e^{ A_n}  =    \e^{- \mathrm{ad}_{A_n}}   O,
$$
and 
  $$
  \alpha_n(O)  :=   \int_0^1 \d s \,   \e^{- s A_n}   O     \e^{ s A_n}  =    \int_0^1 \d s \,   \e^{-s \mathrm{ad}_{A_n}}   O,
  $$
  we get indeed
\begin{align}
H_{n+1} & =  \gamma_n(H_n) + \gamma_n(\nu N)-\nu N  \\[2mm]
 & =  \gamma_n(H_n) - \alpha_n([A_n,\nu N]) \\[2mm]
&=      \gamma_n(D_n)  +   ( \gamma_n(V_n)-V_n)  +  (V_n - [A_n,\nu N])    -  ( \alpha_n([A_n,\nu N])  -[A_n,\nu N])   \\[2mm]
& =      \gamma_n(D_n)  +   ( \gamma_n(V_n)-V_n)  -  ( \alpha_n(V_n)  -V_n).
\end{align}
This equation is identical to the one for the time-dependent setup and all the bounds can be copied.

\section*{Acknowledgements.} 
D.A.\@ acknowledges support by the Alfred Sloan Foundation.  
W.D.R.\@ thanks the Deutsche Forschungsgesellschaft, DFG No. RO 4522/1-1 and the Belgian Interuniversity Attraction Pole (P07/18 Dygest) for financial support. 
Both F.H.\@ and W.D.R.\@ acknowledge the financial support of the ANR grant JCJC and the CNRS InPhyNiTi Grant (MaBoLo). 
\bibliographystyle{plain}
\bibliography{loclibrary}

\end{document}